\documentclass{article}[12pt]

\usepackage{amscd,amsmath,amsthm,amsfonts,amssymb}
\usepackage{graphicx}
\usepackage{multicol}
\usepackage{color}

\def\[{\begin{equation}}
\def\]{\end{equation}}

\begin{document}
\title{\bf A normal form for Hamiltonian-Hopf bifurcations in generalized nonlinear Schr\"odinger equations}
\author{Jianke Yang \\
Department of Mathematics and Statistics \\
University of Vermont \\ Burlington, VT 05401, USA}
\date{}
\maketitle

\begin{abstract}
A normal form is derived for Hamiltonian-Hopf bifurcations of
solitary waves in generalized nonlinear Schr\"odinger equations.
This normal form is a simple second-order nonlinear ordinary
differential equation that is asymptotically accurate in describing
solution dynamics near Hamiltonian-Hopf bifurcations. When the
nonlinear coefficient in this normal form is complex, which occurs
if the second harmonic of the Hopf bifurcation frequency falls
inside the continuous spectrum of the system, the solution of this
normal form will blow up to infinity in finite time, meaning that
solution oscillations near Hamiltonian-Hopf bifurcations will
strongly amplify and eventually get destroyed. When the nonlinear
coefficient of the normal form is real, the normal form can admit
periodic solutions, which correspond to long-lasting solution
oscillations in the original PDE system. Quantitative comparison
between the normal form's predictions and true PDE solutions is also
made in several numerical examples, and good agreement is obtained.

\end{abstract}

\section{Introduction}
Bifurcations of stationary waves are common phenomena in both
conservative and dissipative nonlinear wave systems (see
\cite{pattern_formation,Hopf_dissipative_1976,period_doubling,Akhmediev_pitchfork_1995,Champneys_1996,Akylas_1997,Benard_convection,Faraday,
Panos_2005_pitchfork,Kapitula_2006,Weinstein_2008,Panos_2009,Weinstein_2010,Kirr_2011,Goodman_2011,snaking_2012,Peli_2012,Yang_noswitching_2012,Yang_classification_2012,Yang_PhysicaD_2013}
for instance). Notable examples include symmetry-breaking
bifurcations, fold bifurcations, Hopf bifurcations and
period-doubling bifurcations, all of which have counterparts in
dynamical systems. Bifurcations of stationary waves induce
qualitative changes to the wave behavior and can be used to control
system outcome, thus their studies are both mathematically and
physically important.

Hamiltonian-Hopf bifurcations occur in conservative wave systems,
where pairs of imaginary eigenvalues in the linear-stability
spectrum of stationary waves coalesce and then move off the
imaginary axis, creating oscillatory instability. These linear
instabilities have been well analyzed, and it has been shown that
only collisions of imaginary eigenvalues with opposite Krein
signatures can induce such bifurcations
\cite{MacKay,Peli_book,Kapitula_book}. However, nonlinear wave
dynamics near such bifurcations is less known. One step in this
direction was made by Goodman \cite{Goodman_2011}, where
Hamiltonian-Hopf bifurcations of solitary waves were examined in the
nonlinear Schr\"odinger (NLS) equation with potentials of primarily
symmetric triple-well type. Projecting the solution to three linear
eigenmodes of the potential and making a Galerkin truncation and
symmetry reduction, a Hamiltonian system of ordinary differential
equations (ODEs) for two complex variables was derived. Numerical
simulations of this ODE model showed oscillatory as well as chaotic
solutions at different power levels, which resemble dynamics in the
original PDE system. However, this analysis is only suitable to
Hamiltonian-Hopf bifurcations at low powers of solitons, where the
solution projection onto linear modes is justified (at higher
soliton powers, the error of Galerkin truncation will be
significant). In addition, the derived ODE model is complicated,
which hinders analytical predictions of ODE dynamics. Furthermore,
the resonant effect of higher harmonics of solution oscillations
with the continuous spectrum is invisible in this analysis.

In this paper, we derive a normal form for general Hamiltonian-Hopf
bifurcations of solitary waves (solitons) in NLS equations with
arbitrary external potentials. This normal form is a simple
second-order nonlinear ODE. It is derived near a Hamiltonian-Hopf
bifurcation point by multi-scale perturbation methods and is
asymptotically accurate in describing solution dynamics near
Hamiltonian-Hopf bifurcations. When the nonlinear coefficient in
this normal form is complex, which occurs if the second harmonic of
the Hopf bifurcation frequency is resonant with the continuous
spectrum of the system, the solution of this normal form will blow
up to infinity in finite time, meaning that solution oscillations
near Hamiltonian-Hopf bifurcations will strongly amplify and
eventually get destroyed. When the nonlinear coefficient of the
normal form is real, the normal form can admit periodic solutions,
which correspond to long-lasting solution oscillations in the
original PDE system. Quantitative comparison between the normal
form's predictions and true PDE solutions is also made in several
numerical examples, and good agreement is obtained.

\section{Analytical conditions for Hamiltonian-Hopf bifurcations} \label{s:HHcondition}

We consider the NLS equation with a general external potential,
\[  \label{e:NLS}
iU_t+U_{xx}-V(x)U+\sigma |U|^2U=0,
\]
where $V(x)$ is a real-valued localized potential function, and
$\sigma=\pm 1$ is the sign of nonlinearity. This equation models
nonlinear light propagation in a Kerr medium under paraxial
approximation, as well as dynamics of Bose-Einstein condensates
under mean-field approximation \cite{Kivshar_book,BEC,Yang_SIAM}.
Eq. (\ref{e:NLS}) is a Hamiltonian system. Note that the cubic
nonlinearity and localized linear potential in this model are chosen
primarily for convenience, as the analysis to be developed in this
article can be readily extended to arbitrary forms of nonlinearities
and potentials \cite{Yang_classification_2012}.

Solitons in Eq. (\ref{e:NLS}) have the form
\[  \label{e:soliton}
U(x,t)=e^{i\mu t}u(x),
\]
where $u(x)$ is a real-valued localized function solving the
equation
\[ \label{e:u}
u_{xx}-\mu u-V(x)u+\sigma u^3=0,
\]
and $\mu$ is a real-valued propagation constant. These solitons
exist as continuous families parameterized by $\mu$. Since the
potential $V(x)$ is localized, $\mu$ is positive for all these
solitons.

Linear stability of these solitons is determined by substituting the
normal-mode perturbation
\[
U(x,t)=e^{i\mu t}\left[u(x)+f_1(x)e^{\lambda t}+f_2^*(x)e^{\lambda^*t}\right], \quad f_1, f_2\ll 1
\]
into Eq. (\ref{e:NLS}), which leads to the following linear
eigenvalue problem
\[
iL\left[\begin{array}{c} f_1 \\ f_2 \end{array}\right]=\lambda \left[\begin{array}{c} f_1 \\ f_2 \end{array}\right],
\]
where
\[
L=\left[\begin{array}{cc} \partial_{xx}-\mu-V+2\sigma u^2  & \sigma u^2  \\
-\sigma u^2 & -(\partial_{xx}-\mu-V+2\sigma u^2) \end{array}\right],
\]
$\lambda$ is the eigenvalue, and the superscript `*' represents
complex conjugation. Notice that $\sigma_3L$ is a self-adjoint
operator, where $\sigma_3=\mbox{diag}(1, -1)$ is the third
Pauli-spin matrix. That is, $(\sigma_3L)^A=\sigma_3L$, with the
superscript `\emph{A}' representing the adjoint of an operator. Thus
\[  \label{e:LA}
L^A=\sigma_3 L \sigma_3.
\]

It is easy to see that if $\lambda$ is an eigenvalue of $iL$, so are
$\lambda^*$, $-\lambda$ and $-\lambda^*$. Thus purely-real and
purely-imaginary eigenvalues appear as $\pm \lambda$ pairs, and
other eigenvalues appear as quadruples. If all eigenvalues $\lambda$
lie on the imaginary axis, then the soliton (\ref{e:soliton}) is
linearly stable; otherwise it is linearly unstable.

Hamiltonian-Hopf bifurcations of solitons occur when pairs of
eigenvalues on the imaginary axis collide and move off the imaginary
axis, creating linear oscillatory instability. Suppose this
Hamiltonian-Hopf bifurcation occurs at the soliton with
$\mu=\mu_0>0$, eigenvalues collide at $\lambda=\pm i\omega$ on the
imaginary axis (with $\omega>0$ being the Hopf frequency), and
$i\omega$ is a double eigenvalue of $iL_0$, where $L_0\equiv
L|_{\mu=\mu_0}$. Then the condition for this bifurcation is that the
geometric multiplicity of $i\omega$ is one (less than its algebraic
multiplicity two). In other words, the double eigenvalue $i\omega$
of $iL_0$ admits a single eigenfunction and a generalized
eigenfunction. More explicitly, there exist a single real
eigenfunction $[\psi_1, \psi_2]^T$ and a generalized real
eigenfunction $[\phi_1, \phi_2]^T$ such that
\[
L_0\left[\begin{array}{c} \psi_1 \\ \psi_2 \end{array}\right]=\omega \left[\begin{array}{c} \psi_1 \\ \psi_2 \end{array}\right],
\]
\[  \label{e:geigen}
(L_0-\omega)\left[\begin{array}{c} \phi_1 \\ \phi_2 \end{array}\right]=\left[\begin{array}{c} \psi_1 \\ \psi_2 \end{array}\right],
\]
and no higher generalized eigenfunctions exist, i.e., the equation
\[  \label{e:geigen2}
(L_0-\omega)\left[\begin{array}{c} \chi_1 \\ \chi_2 \end{array}\right]=\left[\begin{array}{c} \phi_1 \\ \phi_2 \end{array}\right]
\]
admits no solutions. Here the superscript `\emph{T}' represents
transpose of a vector. In view of Eq. (\ref{e:LA}), we see that
\[ \label{e:eigenA}
(L_0-\omega)^A \left[\begin{array}{c} \psi_1 \\ -\psi_2 \end{array}\right]=0,
\]
i.e., $[\psi_1, -\psi_2]^T$ is in the kernel of $(L_0-\omega)^A$.
Hence in order for the generalized eigenfunction $[\phi_1,
\phi_2]^T$ to exist in Eq. (\ref{e:geigen}), the right hand side of
this linear inhomogeneous equation must be orthogonal to the adjoint
homogeneous solution $[\psi_1, -\psi_2]^T$, i.e.,
\[  \label{e:cond1}
\int_{-\infty}^{\infty} (\psi_1^2-\psi_2^2) dx=0;
\]
and in order for higher generalized eigenfunctions not to exist in
Eq. (\ref{e:geigen2}), the right hand side of (\ref{e:geigen2}) must
not be orthogonal to $[\psi_1, -\psi_2]^T$, i.e.,
\[  \label{e:cond2}
\int_{-\infty}^{\infty} (\psi_1\phi_1-\psi_2\phi_2) dx \ne 0.
\]

Necessary conditions for Hamiltonian-Hopf bifurcations can also be
formulated using Krein signatures of purely imaginary eigenvalues in
the linear-stability operator $iL$ \cite{MacKay,Kapitula_book}. For
a simple purely imaginary eigenvalue $\lambda=i\omega$ with real
eigenfunction $F=[f_1, f_2]^T$, its Krein signature can be defined
as
\[ \label{e:Krein}
K_\lambda= \mbox{sgn} \langle -\sigma_3LF, F\rangle=\mbox{sgn}\int_{-\infty}^\infty \omega(f_2^2-f_1^2)dx,
\]
where the inner product between two vector functions $f(x)$ and
$g(x)$ is $\langle f, g\rangle=\int_{-\infty}^\infty
f^{*T}\hspace{-0.06cm}g \hspace{0.04cm} dx$. When two such
eigenvalues collide on the imaginary axis, a necessary condition for
Hamiltonian-Hopf bifurcations is that they have opposite Krein
signatures. This necessary condition on Krein signatures is related
to the conditions on eigenfunctions given above. Indeed, under
conditions (\ref{e:cond1})-(\ref{e:cond2}) for eigenfunctions, we
can show that the purely imaginary eigenvalues before collision have
opposite Krein signatures (details are omitted).

Regarding operator $L_0$, it should be added that zero is its
discrete eigenvalue since
\[  \label{e:L0zero}
L_0 \left[\begin{array}{c} u_0 \\ -u_0 \end{array}\right]=0.
\]
In addition,
\[ \label{e:L0Azero}
L_0^A \left[\begin{array}{c} u_0 \\ u_0 \end{array}\right]=0
\]
in view of Eq. (\ref{e:LA}). Furthermore, by differentiating the soliton equation (\ref{e:u}) with respect to $\mu$, we see that
\[  \label{e:L0dmu}
L_0 \left[\begin{array}{c} u_{\mu 0} \\ u_{\mu 0} \end{array}\right]=\left[\begin{array}{c} u_0 \\ -u_0 \end{array}\right],
\]
where $u_{\mu 0}(x) \equiv u_\mu(x; \mu)|_{\mu=\mu_0}$. These relations will be used in later analysis.

In the next section, we will investigate solution dynamics near
Hamiltonian-Hopf bifurcations. For that purpose, we make the
following additional assumptions:

\begin{enumerate}
\item $2i\omega$ is not a discrete eigenvalue of $iL_0$;
\item Defining the linearization operator of the soliton equation
(\ref{e:u}) at $\mu=\mu_0$ as
\[
M=\partial_{xx}-\mu_0-V(x)+3\sigma u_0^2,
\]
we assume that the kernel of $M$ is empty;
\item Defining the power of solitons $u(x; \mu)$ as
\[
P(\mu)=\int_{-\infty}^\infty u^2(x; \mu) dx,
\]
then we assume that $P'(\mu_0) \ne 0$.
\end{enumerate}
The first assumption forbids nonlinearity-induced second-harmonic
resonance of Hopf-bifurcation eigenmodes; the second assumption
prohibits other potential bifurcations of solitons at this
Hamiltonian-Hopf bifurcation point \cite{Yang_classification_2012};
and the third assumption excludes additional linear instabilities
\cite{Yang_SIAM}. These assumptions will be needed for our analysis
to carry through.

\section{A normal form for Hamiltonian-Hopf bifurcations}

In this section, we derive an asymptotically accurate ODE model (a
normal form) for wave dynamics near Hamiltonian-Hopf bifurcations.
This normal form uses only information at the bifurcation point.

Near a Hamiltonian-Hopf bifurcation point $\mu=\mu_0$, the full PDE
solution can be expanded into the following perturbation series,
\[ \label{e:Uexpansion}
U(x, t)=e^{i\theta} \left[u_0(x)+\epsilon U_1(x,t,T) +\epsilon^2 U_2(x,t,T) +\dots\right],
\]
where
\[
\theta(t, T)=\mu_0t +\epsilon \int \mu_1(T) \hspace{0.05cm} dT +\epsilon^2 \int\mu_2(T) \hspace{0.05cm} dT +\dots,
\]
$T=\epsilon t$, and $0<\epsilon \ll 1$. Substituting this expansion
into Eq. (\ref{e:NLS}), the $O(1)$ equation is automatically
satisfied. At $O(\epsilon)$, the equation for $U_1$ is found to be
\[
(i\partial_t+\partial_{xx}-\mu_0-V+2\sigma u_0^2)U_1+\sigma u_0^2 U_1^*=0.
\]
This equation can be rewritten as
\[
\left(i\partial_t+L_0\right) \left[\begin{array}{c} U_1 \\ U_1^* \end{array}\right]=0.
\]
In view of the conditions for Hamiltonian-Hopf bifurcations in Sec.
\ref{s:HHcondition}, $i\omega$ is a double eigenvalue of $iL_0$ with
a single real eigenfunction $[\psi_1, \psi_2]^T$. Due to eigenvalue
symmetry, $-i\omega$ is also a double eigenvalue of $iL_0$ with a
single real eigenfunction $[\psi_2, \psi_1]^T$. Thus the non-secular
localized solution for $U_1$ is a slowly modulated Hopf oscillation
mode
\[ \label{s:U1}
U_1=B(T)\psi_1(x)e^{i\omega t}+B^*(T)\psi_2(x)e^{-i\omega t},
\]
where $B(T)$ is a complex envelope function to be determined.

At $O(\epsilon^2)$, the equation for $U_2$ is
\begin{eqnarray}
&& \hspace{0cm} (i\partial_t+\partial_{xx}-\mu_0-V+2\sigma u_0^2)U_2+\sigma u_0^2 U_2^* \nonumber \\ && \hspace{-0.1cm}  =\mu_1u_0 -iU_{1T}  -\sigma u_0\left(2|U_1|^2+U_1^2\right).
\end{eqnarray}
When the $U_1$-formula (\ref{s:U1}) is utilized, this $U_2$ equation becomes
\begin{eqnarray}
&& \hspace{-1.2cm} (i\partial_t+\partial_{xx}-\mu_0-V+2\sigma u_0^2)U_2+\sigma u_0^2 U_2^*=\mu_1u_0   \nonumber \\
&& \hspace{-1.2cm} -2\sigma
|B|^2u_0(\psi_1^2+\psi_1\psi_2+\psi_2^2)  -iB_T\psi_1e^{i\omega t}-iB_T^*\psi_2e^{-i\omega t} \nonumber \\
&& \hspace{-1.2cm} -\sigma B^2u_0(2\psi_1\psi_2+\psi_1^2)e^{2i\omega t}-\sigma
B^{*2}u_0(2\psi_1\psi_2+\psi_2^2)e^{-2i\omega t}.
\end{eqnarray}
In view of Eqs. (\ref{e:geigen}) and (\ref{e:L0dmu}), the solution to this $U_2$ equation is
\begin{eqnarray} \label{s:U2}
&& \hspace{-0.5cm} U_2= \mu_1 u_{\mu 0}-\sigma |B|^2h -iB_T\phi_1 e^{i\omega t}+iB_T^*\phi_2 e^{-i\omega t}  \nonumber \\
&& \hspace{0.2cm} -\sigma B^2g_1 e^{2i\omega t}-\sigma B^{*2} g_2^* e^{-2i\omega t},
\end{eqnarray}
where $h(x)$ is a real localized function defined by
\[  \label{e:h}
h=M^{-1}\left[2u_0(\psi_1^2+\psi_1\psi_2+\psi_2^2)\right],
\]
$[\phi_1(x), \phi_2(x)]^T$ is the real generalized eigenfunction
defined in (\ref{e:geigen}), and $[g_1(x), g_2(x)]^T$ solves the
equation
\[  \label{e:g12}
(L_0-2\omega)\left[\begin{array}{c} g_1 \\ g_2 \end{array}\right]=
\left[\begin{array}{c} u_0(2\psi_1\psi_2+\psi_1^2) \\  -u_0(2\psi_1\psi_2+\psi_2^2)\end{array}\right].
\]
Note that the kernel of $M$ is empty due to our second assumption in
the end of Sec. \ref{s:HHcondition}, thus a localized real function $h(x)$ as defined
in Eq. (\ref{e:h}) exists and is unique.

An important remark is in order regarding the nature of the solution
$[g_1, g_2]^T$ to Eq. (\ref{e:g12}), and this hinges on whether $2i\omega$ lies inside the continuous
spectrum of the linear-stability operator $iL_0$. Recall that the potential $V(x)$ is localized. Then for the soliton $u_0(x)$, the continuous spectrum of $iL_0$ is $i(-\infty, -\mu_0]\bigcup i[\mu_0, +\infty)$ on the imaginary axis.

(1) If $2i\omega$ does not lie in this continuous spectrum, meaning that
$2\omega < \mu_0$,  then in view of our first
assumption in the end of Sec. \ref{s:HHcondition}, the kernel of $L_0-2\omega$ is
empty, thus a localized real solution $[g_1, g_2]^T$ to Eq.
(\ref{e:g12}) exists and is unique.

(2) If $2i\omega$ lies inside the continuous spectrum of $iL_0$,
i.e., $2\omega > \mu_0$, then resonance with the continuous spectrum
occurs. In this case, the forcing term on the right side of Eq.
(\ref{e:g12}) will excite continuous-wave radiation, which appears
in the $g_2$ component. This radiation must spread from the central
region to the far field, i.e., it must satisfy the Sommerfeld
radiation condition
\begin{equation}
\left[\begin{array}{c} g_1 \\ g_2 \end{array}\right] \to \left\{ \begin{array}{ll}  \left[\begin{array}{c} 0
\\ R_+ e^{-ikx} \end{array}\right], & x\gg 1, \\  \left[\begin{array}{c} 0 \\ R_- e^{ikx} \end{array}\right],
& x\ll -1,\end{array} \right.
\label{e:BC}
\end{equation}
where $k = \sqrt{2\omega-\mu_0}$ is the wavenumber of large-$x$ radiation with frequency $2\omega$, and $R_{\pm}$ are constants
which measure the radiation amplitudes at large-$x$.
A consequence of the Sommerfeld radiation condition is that the
resulting solution $[g_1, g_2]^T$ is complex and nonlocal, and this solution can be uniquely determined by various methods \cite{PeliYang2002,Yang_SIAM}.

After the $U_2$ solution (\ref{s:U2}) is obtained, we now proceed to the $U_3$ equation at
order $\epsilon^3$. This $U_3$ equation is
\begin{eqnarray}
&& \hspace{-0.85cm} (i\partial_t+\partial_{xx}-\mu_0-V+2\sigma u_0^2)U_3+\sigma u_0^2 U_3^*  \nonumber \\
&& \hspace{-0.95cm}   =\mu_1U_1+\mu_2u_0-iU_{2T} \nonumber \\
&& \hspace{-0.8cm}   -\sigma\left( 2u_0U_1U_2+ 2u_0U_1^*U_2+2u_0U_1U_2^*+|U_1|^2U_1\right).
\end{eqnarray}
Inserting the $U_1$ and $U_2$ formulae (\ref{s:U1}) and (\ref{s:U2}), this
$U_3$ equation reduces to
\begin{eqnarray}
&& \hspace{-0.5cm} (i\partial_t+\partial_{xx}-\mu_0-V+2\sigma u_0^2)U_3+\sigma u_0^2 U_3^*  \nonumber \\
&& \hspace{-0.55cm}   = Q_0 + Q_1 e^{i\omega t} + Q_2^*e^{-i\omega t} +Q_3 e^{2i\omega t} + Q_4^* e^{-2i\omega t} \nonumber \\
&& \hspace{-0.4cm} +Q_5 e^{3i\omega t} + Q_6^* e^{-3i\omega t},
\end{eqnarray}
where
\begin{eqnarray}
&& \hspace{-0.5cm} Q_0=-i\left[\mu_{1T}u_{\mu 0}-\sigma h (|B|^2)_T \right] \nonumber \\
&& \hspace{0.3cm}  -2\sigma u_0\left[iBB_T^*(\psi_1\phi_1+\psi_2\phi_2 +\psi_1\phi_2) \right. \nonumber \\
&& \hspace{1.3cm}  \left. -iB^*B_T(\psi_1\phi_1+\psi_2\phi_2 +\psi_2\phi_1)\right],
\end{eqnarray}
\begin{eqnarray}
&& \hspace{-0.5cm} Q_1=\mu_1 B\left[ \psi_1-2\sigma u_0 u_{\mu 0}(2\psi_1+\psi_2)\right]
-B_{TT} \phi_1   \nonumber \\
&& \hspace{0.3cm} -\sigma |B|^2B \left[ \psi_1(2\psi_2^2+\psi_1^2)-2\sigma u_0 h (2\psi_1+\psi_2) \right.
\nonumber \\ && \hspace{1.7cm}
\left. -2\sigma u_0 (\psi_1 g_1 +\psi_2 g_1 +\psi_2 g_2)\right],
\end{eqnarray}
\begin{eqnarray}
&& \hspace{-0.5cm} Q_2=\mu_1 B\left[ \psi_2-2\sigma u_0 u_{\mu 0}(2\psi_2+\psi_1)\right]
+B_{TT} \phi_2   \nonumber \\
&& \hspace{0.3cm} -\sigma |B|^2B \left[ \psi_2(2\psi_1^2+\psi_2^2)-2\sigma u_0 h (2\psi_2+\psi_1) \right.
\nonumber \\ && \hspace{1.7cm}
\left. -2\sigma u_0 (\psi_1 g_1 +\psi_1 g_2 +\psi_2 g_2)\right],
\end{eqnarray}
and $Q_3, Q_4, Q_5, Q_6$ are functions that are unimportant in our analysis and are thus not shown here.
This $U_3$ solution can be decomposed as
\begin{eqnarray}
&& \hspace{-1cm} U_3(x, t)=U_{30}+U_{31} e^{i\omega t} + U_{32}^*e^{-i\omega t} \nonumber \\
&& \hspace{-0.4cm} +U_{33} e^{2i\omega t} + U_{34}^* e^{-2i\omega t} +U_{35} e^{3i\omega t} + U_{36}^* e^{-3i\omega t},
\end{eqnarray}
where
\[
L_0 \left[\begin{array}{c}  U_{30} \\ U_{30}^* \end{array}\right]=\left[\begin{array}{c} Q_0 \\ -Q_0^* \end{array}\right],
\]
and
\[
\left(L_0-\omega\right) \left[\begin{array}{c}  U_{31} \\ U_{32} \end{array}\right]=\left[\begin{array}{c} Q_1 \\ -Q_2 \end{array}\right].
\]
The Fredholm solvability conditions of the above two equations are that their inhomogeneous terms on the right hand sides be orthogonal to the localized adjoint homogeneous solutions. In view of Eqs. (\ref{e:eigenA}) and (\ref{e:L0Azero}), these solvability conditions are
\[
\int_{-\infty}^\infty u_0 (Q_0-Q_0^*) dx=0,
\]
and
\[
\int_{-\infty}^\infty \left(\psi_1 Q_1 + \psi_2 Q_2 \right) dx=0.
\]
Inserting the expressions of $Q_0, Q_1$ and $Q_2$ above, these solvability conditions lead to the following dynamical equations for the slow variables $\mu_1(T)$ and $B(T)$,
\begin{eqnarray}
&& \mu_{1T} = \alpha (|B|^2)_T,    \label{e:mu1T}       \\
&& B_{TT}-\beta \mu_1 B +\gamma |B|^2B=0,   \label{e:BTT}
\end{eqnarray}
where
\[ \label{d:alpha}
\alpha=\frac{2\sigma}{P'(\mu_0)} \int_{-\infty}^\infty u_0 \left[ h-u_0(\psi_1\phi_2 - \psi_2\phi_1)\right] dx,
\]
\[  \label{d:beta}
\beta= \frac{\int_{-\infty}^\infty \left[ \psi_1^2 +\psi_2^2 -4\sigma u_0 u_{\mu 0} (\psi_1^2 + \psi_1 \psi_2 + \psi_2^2 ) \right] dx}
{\int_{-\infty}^{\infty} (\psi_1\phi_1-\psi_2\phi_2) dx},
\]
and
\[ \label{d:gamma}
\gamma = \frac{\sigma \int_{-\infty}^\infty
\left\{   \begin{array}{l} \psi_1^4 +4\psi_1^2\psi_2^2 + \psi_2^4-4\sigma u_0 h (\psi_1^2 + \psi_1 \psi_2 + \psi_2^2 ) \\
-2\sigma u_0 [ \psi_1^2 g_1 +2\psi_1\psi_2 (g_1+g_2)+\psi_2^2g_2 ]  \end{array} \right\} dx}
{\int_{-\infty}^{\infty} (\psi_1\phi_1-\psi_2\phi_2) dx}.
\]
In view of Eq. (\ref{e:cond2}) and our third assumption in the end
of Sec. \ref{s:HHcondition}, the above three constants are well
defined. Dynamical equations (\ref{e:mu1T})-(\ref{e:BTT}) are our
normal form for nonlinear wave dynamics near Hamiltonian-Hopf
bifurcations in the PDE (\ref{e:NLS}).

Of the three constants in this normal form, $\alpha$ and $\beta$ are
always real. But $\gamma$ is real only when $2\omega < \mu_0$ and
becomes complex if $2\omega > \mu_0$, since the functions $(g_1,
g_2)$ involved in the definition of $\gamma$ are real and complex
under those conditions respectively (see earlier text in this
section).

The physical meaning of parameter $\beta$ in Eq. (\ref{e:BTT}) can
be revealed by considering small-$B$ solutions. In this case,
$\mu_{1T}\approx 0$, i.e., $\mu_1$ is approximately a constant, and
the resulting solution (\ref{e:Uexpansion}) is approximately a
soliton with propagation constant $\mu\approx \mu_0+\epsilon^2
\mu_1$. The small perturbation $B$ satisfies $B_{TT} -\beta \mu_1
B\approx 0$, thus $B\sim e^{\tilde{\lambda} t}$, where
$\tilde{\lambda} = \pm \epsilon \sqrt{\beta \mu_1}$. If
$\beta\mu_1<0$, then the perturbation $B$ is bounded, meaning that
the soliton is before the Hamiltonian-Hopf bifurcation. When
$\beta\mu_1>0$, the perturbation $B$ exponentially grows, meaning
that the soliton is after the Hamiltonian-Hopf bifurcation. In the
latter case, recalling Eq. (\ref{s:U1}), we see that the quartet of
complex eigenvalues born out of Hamiltonian-Hopf bifurcations are
$\lambda\approx \pm i\omega \pm \sqrt{\beta (\mu-\mu_0)}$. Thus,
physically $\beta$ determines the growth rate of perturbations after
Hamiltonian-Hopf bifurcations.

The normal form (\ref{e:mu1T})-(\ref{e:BTT}) can be further
simplified. Notice that Eq. (\ref{e:mu1T}) can be integrated once,
and we get
\[
\mu_1=\alpha |B|^2 +c_0,
\]
where $c_0$ is a real constant which can be determined from the
initial conditions of $\mu_1$ and $B$. Substituting this equation
into (\ref{e:BTT}), we obtain a decoupled equation for $B$ as
\[  \label{e:BTT2}
B_{TT}-\widehat{\beta}\hspace{0.04cm} B+\widehat{\gamma}\hspace{0.05cm} |B|^2B=0,
\]
where
\[
\widehat{\beta}=\beta \hspace{0.04cm} c_0, \quad \widehat{\gamma}=\gamma-\alpha\beta.
\]
In this equation, $\widehat{\beta}$ is a real constant, while
$\widehat{\gamma}$ can be real or complex depending on the reality
of $\gamma$. Eq. (\ref{e:BTT2}) is a reduced normal form for
nonlinear wave dynamics near Hamiltonian-Hopf bifurcations. This
reduced normal form is a second-order nonlinear ODE for a complex
variable $B$.

For a different symmetry-breaking bifurcation of solitons in the NLS
equation with external potentials (\ref{e:NLS}), a normal form
similar to (\ref{e:BTT2}) was derived by Pelinovsky and Phan
\cite{Peli_2012}, but the variable and the nonlinear coefficient in
their normal form were both real. For the present Hamiltonian-Hopf
bifurcation, both the variable $B$ and the nonlinear coefficient
$\widehat{\gamma}$ can be complex, and this fact will have important
consequences on the solution dynamics, as we will elaborate below.

\section{Solution behaviors in the normal form}

Now we analyze solution dynamics in the reduced normal form
(\ref{e:BTT2}).

\subsection{Solutions for real nonlinear coefficient $\widehat{\gamma}$}

When $\widehat{\gamma}$ is real, Eq. (\ref{e:BTT2}) can be solved exactly. In polar variables $B=be^{i\xi}$, where $b$ is the amplitude and $\xi$ the phase of $B$, Eq. (\ref{e:BTT2}) yields
\[  \label{e:bTT}
b_{TT}-\widehat{\beta} \hspace{0.04cm} b+\widehat{\gamma} \hspace{0.04cm} b^3-\xi_{T}^2 b=0,
\]
\[
(b^2\xi_T)_T=0.
\]
The second equation shows that
\[
b^2\xi_T=w_0,
\]
where $w_0$ is a real constant. Using this relation, Eq. (\ref{e:bTT}) reduces to
\[
b_{TT}-\widehat{\beta} \hspace{0.04cm} b+\widehat{\gamma} \hspace{0.04cm} b^3-\frac{w_0^2}{b^3}=0,
\]
which describes the motion of a particle in a potential well,
\[
b_{TT} + W'(b)=0,
\]
where the potential $W(b)$ is
\[
W(b)=-\frac{1}{2}\widehat{\beta} \hspace{0.04cm} b^2+\frac{1}{4}\widehat{\gamma} \hspace{0.04cm} b^4+\frac{w_0^2}{2b^2}.
\]
Then depending on system parameters ($\widehat{\beta},
\widehat{\gamma}$) and initial conditions, the solution $B(T)$ can
be completely obtained.

A special but important class of solutions arises when $w_0=0$, i.e., $\xi_T=0$. In view of gauge invariance, these solutions are equivalent to real-$B$ solutions, which will be elaborated further below.

\begin{figure}[h!]
\centerline{\includegraphics[width=0.8\textwidth]{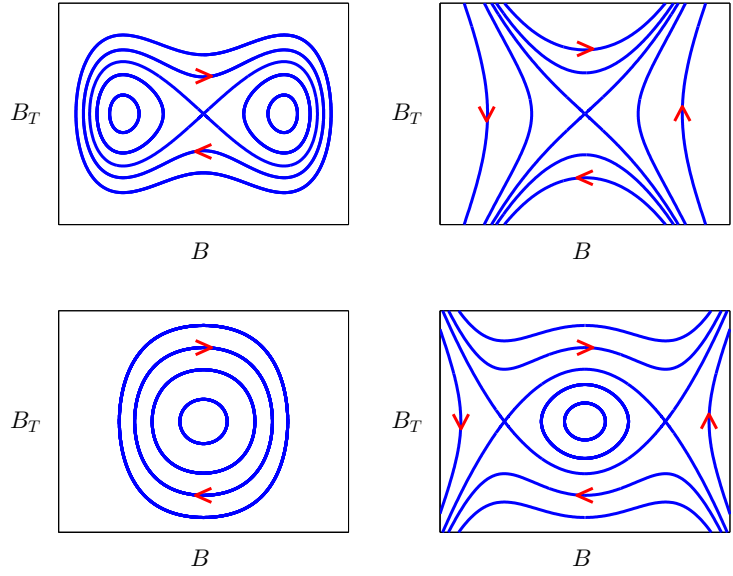}}
\caption{(Color online) Phase portraits of the normal form (\ref{e:BTT2}) for real values of $\widehat{\gamma}$ and $B$. Upper left: $\widehat{\beta}>0, \widehat{\gamma}>0$;
upper right: $\widehat{\beta}>0, \widehat{\gamma}<0$; lower left: $\widehat{\beta}<0, \widehat{\gamma}>0$; lower right: $\widehat{\beta}<0, \widehat{\gamma}<0$.  }
\label{f:phase}
\end{figure}

When $B$ is real, Eq. (\ref{e:BTT2}) becomes
\[ \label{e:BTTreal}
B_{TT}-\widehat{\beta}\hspace{0.04cm} B+\widehat{\gamma}\hspace{0.05cm} B^3=0.
\]
The solution dynamics in this equation is best illustrated by its phase portrait in the $(B, B_T)$ plane.
Depending on the signs of $\widehat{\beta}$ and $\widehat{\gamma}$, four qualitatively different phase portraits are displayed in Fig.~\ref{f:phase}.
These phase portraits show that when $\widehat{\gamma}>0$, the trajectories are all bounded and almost always periodic (see the left panels), which correspond to time-periodic bound states (\ref{e:Uexpansion}) in the PDE system (\ref{e:NLS}). When $\widehat{\gamma}<0$, trajectories all escape to infinity if $\widehat{\beta}>0$ (upper right panel), and escape to infinity for larger initial conditions and stay bounded and periodic for smaller initial conditions if $\widehat{\beta}<0$ (lower right panel).

\subsection{Solutions for complex nonlinear coefficient $\widehat{\gamma}$}

When $\widehat{\gamma}$ is complex, under polar variables $B=be^{i\xi}$ and the notation $w\equiv r^2\xi_T$, Eq. (\ref{e:BTT2}) becomes
\[  \label{e:bTTc}
b_{TT}-\widehat{\beta} \hspace{0.04cm}  b+\mbox{Re}(\widehat{\gamma}) \hspace{0.04cm} b^3-\frac{w^2}{b^3}=0,
\]
\[ \label{e:wT}
w_T=-\mbox{Im}(\widehat{\gamma}) \hspace{0.04cm} b^4.
\]
The first equation shows that $b$ is bounded away from zero. Then the second equation shows that $w$ keeps increasing or decreasing to infinity. Viewing Eq. (\ref{e:bTTc}) as the motion of a particle in a potential well, when $w$ goes to infinity, the potential term $w^2/(2b^2)$ dominates, hence the solution $b$ escapes to infinity as well.

We can further determine precisely in which manner the solution $(b, w)$ escapes to infinity. The nature of nonlinearity in Eqs. (\ref{e:bTTc})-(\ref{e:wT}) indicates that the solution will escape to infinity in finite time. Suppose the time of blowup is $T_0$, and
\[ \label{e:asymbw}
b(T) \sim \frac{b_1}{(T_0-T)^m}, \quad w(T) \sim \frac{w_1}{(T_0-T)^n},  \quad T\sim T_0,
\]
where $b_1, w_1, m, n$ are real constants to be determined.
Substituting these asymptotics into Eq. (\ref{e:wT}), we get
\[
n=4m-1, \quad w_1=-\frac{\mbox{Im}(\widehat{\gamma}) \hspace{0.04cm} b_1^4}{n}.
\]
Substituting the asymptotics (\ref{e:asymbw}) and the above equation into (\ref{e:bTTc}) and using dominant balance, we get
\[
m=1, \quad 2b_1+\mbox{Re}(\widehat{\gamma}) \hspace{0.04cm} b_1^3=\frac{w_1^2}{b_1^3}.
\]
After solving $(b_1, w_1)$ from the above two equations, we find that the blowup profile of the solution $B(T)$ is
\[ \label{e:asymbw2}
b(T) \sim \frac{b_1}{T_0-T}, \quad \xi(T) \sim c_1\mbox{ln}(T_0-T),  \quad T\sim T_0,
\]
where
\[
b_1=\frac{3}{\sqrt{2}\: \mbox{Im}(\widehat{\gamma})}
\sqrt{\mbox{Re}(\widehat{\gamma})+\sqrt{\left[\mbox{Re}(\widehat{\gamma})\right]^2+\frac{8}{9}\left[\mbox{Im}(\widehat{\gamma})\right]^2}},
\]
and
\[
c_1=\frac{1}{3} \mbox{Im}(\gamma) \hspace{0.04cm} b_1^2.
\]
Notice that this collapsing profile is completely determined by the
nonlinear coefficient $\gamma$, except for the collapsing time
$T_0$. Notice also that both the amplitude $b$ and phase $\xi$
collapse, but at different rates.

\begin{figure}[h!]
\centerline{\includegraphics[width=0.8\textwidth]{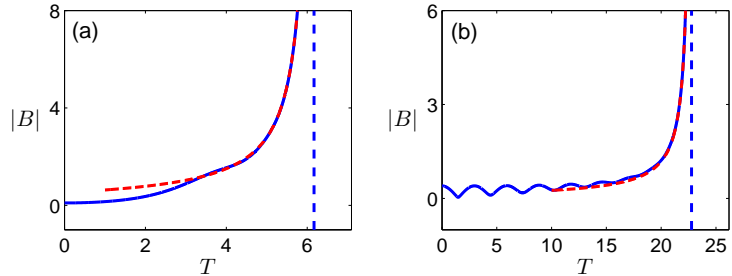}}
\caption{(Color online) Collapsing solutions of the normal form (\ref{e:BTT2}) when $\widehat{\gamma}$ is complex.
(a) $\widehat{\beta}=1$ and $\widehat{\gamma}=1+i$; (b) $\widehat{\beta}=-1$ and $\widehat{\gamma}=1+i$.
Solid blue lines are numerically obtained solutions of the normal form; dashed red lines are analytical collapsing profiles (\ref{e:asymbw2});
and vertical dashed lines are collapsing times. } \label{f:collapse}
\end{figure}

To illustrate these collapsing solutions for complex
$\widehat{\gamma}$, we take $\widehat{\gamma}=1+i$ and two different
values of $\pm 1$ for $\widehat{\beta}$. For these parameter
choices, we have computed Eq. ({\ref{e:BTT2}) for various initial
conditions and found that the solution always collapses, and the
collapsing profile exactly matches that predicted by Eq.
(\ref{e:asymbw2}). Two examples of such computations are displayed
in Fig.~\ref{f:collapse}, along with analytical collapsing profiles
for comparison. For these two values of $\widehat{\beta}$, the two
solutions $B(T)$ initially behave quite differently. But they both
approach the same collapsing profile (\ref{e:asymbw2}) in the end.

Collapsing solutions of $B(T)$ correspond to PDE solutions
(\ref{e:Uexpansion}) where the Hopf oscillation mode (\ref{s:U1})
strongly intensifies and the underlying soliton $u_0(x)$ eventually
breaks up. This solution collapse occurs when $\widehat{\gamma}$ is
complex, i.e., when $2\omega > \mu_0$, where resonance with the
continuous spectrum arises in the $U_2$ solution (\ref{s:U2}). Due
to this resonance, energy is channeled from the underlying soliton
to the Hopf oscillation mode (\ref{s:U1}) and continuous-wave
radiation. When $2\omega < \mu_0$, this resonance does not occur in
$U_2$, $\widehat{\gamma}$ is real, and the ODE solution $B(T)$ can
be bounded (see Fig.~\ref{f:phase}). However, due to
nonlinearity-induced higher harmonics, there always exists a higher
harmonic $e^{in\omega}$ for some integer $n$ so that $n\omega >
\mu_0$, in which case resonance with the continuous spectrum will
occur in the $U_{n}$ solution of the perturbation expansion
(\ref{e:Uexpansion}). As a consequence, when $2\omega < \mu_0$, the
$B$ solution is still expected to collapse, and the underlying
soliton $u_0(x)$ is still expected to break up, except that these
events will take much longer time to develop since the resonance is
at higher orders of the perturbation expansion.

\section{Numerical examples}

In this section, we use several numerical examples to illustrate the
theory and compare the numerical results with the normal-form's
predictions.

\textbf{Example 1}\ In our first example, we take
\[ \label{e:Vexample1}
V(x)=-3\left[\mbox{sech}^2(x+1)+\mbox{sech}^2(x-1)\right],
\]
which is a symmetric double-well potential, and $\sigma=1$ (focusing nonlinearity). This potential is displayed in Fig.~\ref{f:example1}(a).
This potential admits three linear discrete eigenvalues $\mu_a < \mu_b < \mu_c$, with the middle one $\mu_b \approx 1.4104$. From this linear eigenmode, a family of dipole-type solitons bifurcates out. The reason for choosing this dipole-soliton family bifurcated from the middle linear eigenmode is that, at low amplitudes of these solitons, it can be readily shown that the linear-stability operator $iL$ has two purely imaginary eigenvalues of opposite Krein signatures in the upper half plane. Specifically, one imaginary eigenvalue is approximately $i(\mu_c-\mu_b)$ and it has negative Krein signature, and the other imaginary eigenvalue is approximately $i(\mu_b-\mu_a)$ and it has positive Krein signature.
The presence of these two imaginary eigenvalues of opposite Krein signatures makes Hamiltonian-Hopf bifurcation possible as the amplitude of the soliton increases (see Sec.~\ref{s:HHcondition}).

In Fig.~\ref{f:example1}(d) the power curve of this soliton family
is displayed. At two marked points of this power curve where
$\mu=\mu_0\pm 0.05$, profiles of the solitons are plotted in
Fig.~\ref{f:example1}(b,c). Linear-stability spectra of these
solitons are shown in Fig.~\ref{f:example1}(e,f). It is seen that
for the lower-power soliton [see panel (b)], the spectrum is
all-imaginary with a pair of discrete imaginary eigenvalues close to
each other in the upper half plane [see panel (e)]. This pair of
imaginary eigenvalues originate from the eigenvalues
$i(\mu_c-\mu_b)$ and $i(\mu_b-\mu_a)$ of zero-power solitons and
thus have opposite Krein signatures. For the higher-power soliton
[see panel (c)], this pair of discrete eigenvalues have collided and
bifurcated off the imaginary axis, indicating that a
Hamiltonian-Hopf bifurcation has occurred [see panel (f)]. The exact
location of this Hamiltonian-Hopf bifurcation is at $\mu_0\approx
1.8572$, where the two discrete eigenvalues coalesce at $i\omega$,
with $\omega\approx 1.3450$. Notice that $2\omega> \mu_0$, thus
$\gamma$ is complex. Numerical values for constants $\alpha, \beta$
and $\gamma$ in the normal form (\ref{e:mu1T})-(\ref{e:BTT}) are
found to be
\[ \label{v:alphagamma}
\alpha\approx 0.2660, \quad \beta\approx  0.0476, \quad \gamma\approx 0.0780 + 0.0144i.
\]
In obtaining these numbers, the eigenfunction $[\psi_1, \psi_2]^T$ is normalized so that the maximum of $\psi_1^2+\psi_2^2$ is one.

\begin{figure}[h!]
\centerline{\includegraphics[width=0.8\textwidth]{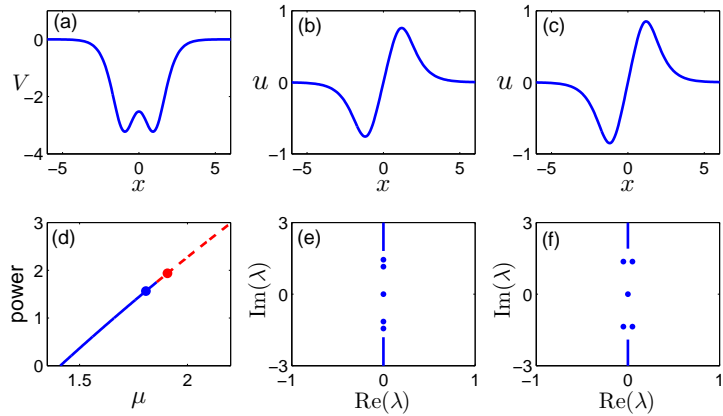}}
\caption{(Color online) Solitons and their stability spectra in Example 1. (a) the double-well potential (\ref{e:Vexample1});
(b, c) profiles of dipole solitons at the marked blue and red points of the power curve in (d) respectively; blue color of the power curve represents stable solitons, and red color represents unstable solitons; (e, f) linear-stability spectra of the solitons in (b,c) respectively.  } \label{f:example1}
\end{figure}

Now we numerically investigate the dynamics of solutions near this
Hamiltonian-Hopf bifurcation point. First we consider dynamics above
the bifurcation. For this purpose, we take the initial condition of
the PDE as a perturbation of the unstable soliton in
Fig.~\ref{f:example1}(c) by Hopf oscillation mode (\ref{s:U1}),
corresponding to $\epsilon^2\mu_1|_{t=0}=0.05$ and $\epsilon
B|_{t=0}=0.01$ in the perturbation solution (\ref{e:Uexpansion}).
Whole-field evolution of this initial condition is illustrated in
Fig.~\ref{f:evolution1}(a), and the amplitude evolution at $x=1.2$
is shown Fig.~\ref{f:evolution1}(b). It is seen that the underlying
soliton breaks up and the solution evolves into oscillations which
grow stronger over time. Taking the initial condition of the normal
form (\ref{e:BTT2}) corresponding to the above initial condition of
the PDE, the solution $B$ of the normal form is plotted in
Fig.~\ref{f:evolution1}(c). This $B$ solution grows and eventually
blows up in finite time (the blow-up part is not shown since the
perturbation theory will be invalid when the blow-up occurs). Using
this $B$ solution, we have reconstructed the perturbation solution
(\ref{e:Uexpansion}) (up to order $\epsilon^2$), and the amplitude
evolution at $x=1.2$ is shown Fig.~\ref{f:evolution1}(d). Comparing
this analytically reconstructed amplitude evolution with the
numerical one in panel (b), we can see that the normal form
(\ref{e:BTT2}) gives a good qualitative and quantitative prediction
of the PDE dynamics above Hamiltonian-Hopf bifurcations over a long
period of time.

\begin{figure}[h!]
\centerline{\includegraphics[width=0.8\textwidth]{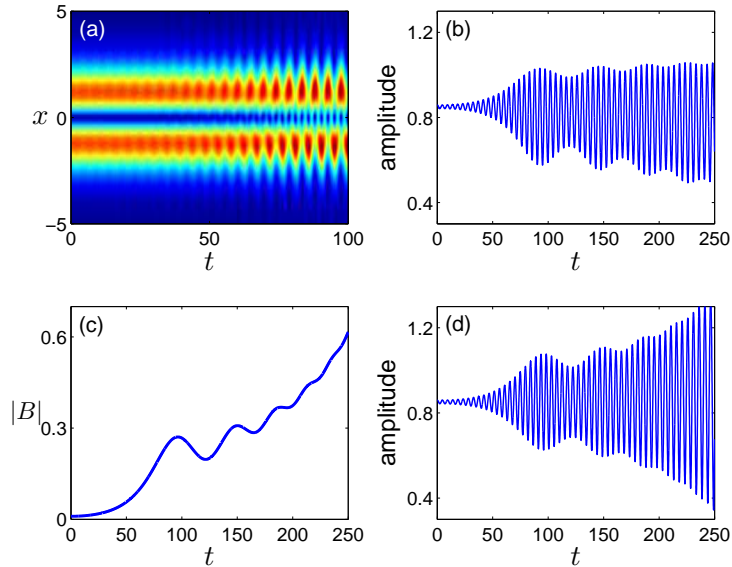}}
\caption{(Color online) Solution evolution above Hamiltonian-Hopf bifurcation in Example 1. (a) Contour plot of the PDE solution in the $(x,t)$ plane; (b) amplitude evolution of the PDE solution versus time. Here the amplitude is measured as $|U(x,t)|$ at location $x=1.2$;
(c) the solution $|B|$ obtained from the normal form (\ref{e:BTT2}); (d) amplitude evolution of the analytically reconstructed perturbation solution (\ref{e:Uexpansion}).
}  \label{f:evolution1}
\end{figure}

Next we consider solution dynamics in Example 1 below the
Hamiltonian-Hopf bifurcation. For this purpose, we take the initial
condition of the PDE as a perturbation of the stable soliton in
Fig.~\ref{f:example1}(b) by Hopf oscillation mode (\ref{s:U1}),
corresponding to $\epsilon^2\mu_1|_{t=0}=-0.05$ and $\epsilon
B|_{t=0}=0.08$ in the perturbation solution (\ref{e:Uexpansion}).
Evolution of this initial condition is illustrated in
Fig.~\ref{f:evolution2}(a,b,c). It is seen that the soliton
eventually also breaks up due to growing oscillations. This
instability is interesting since the underlying soliton is linearly
stable [see Fig.~\ref{f:example1}(e)]. Thus this instability is a
nonlinear instability. The analytical reason for this nonlinear
instability is that the nonlinear coefficient $\widehat{\gamma}$ in
the normal form (\ref{e:BTT2}) is complex [see Eq.
(\ref{v:alphagamma})], thus the ODE solution $B$ always collapses
both below and above the Hamiltonian-Hopf bifurcation (see
Fig.~\ref{f:collapse}). From the ODE solution of the normal form
(\ref{e:BTT2}), we have also reconstructed the perturbation solution
(\ref{e:Uexpansion}), and the amplitude evolution at $x=1.2$ is
shown Fig.~\ref{f:evolution2}(d). In this case, the normal form also
gives a good qualitative and quantitative prediction of the PDE
dynamics over long times.

\begin{figure}[h!]
\centerline{\includegraphics[width=0.8\textwidth]{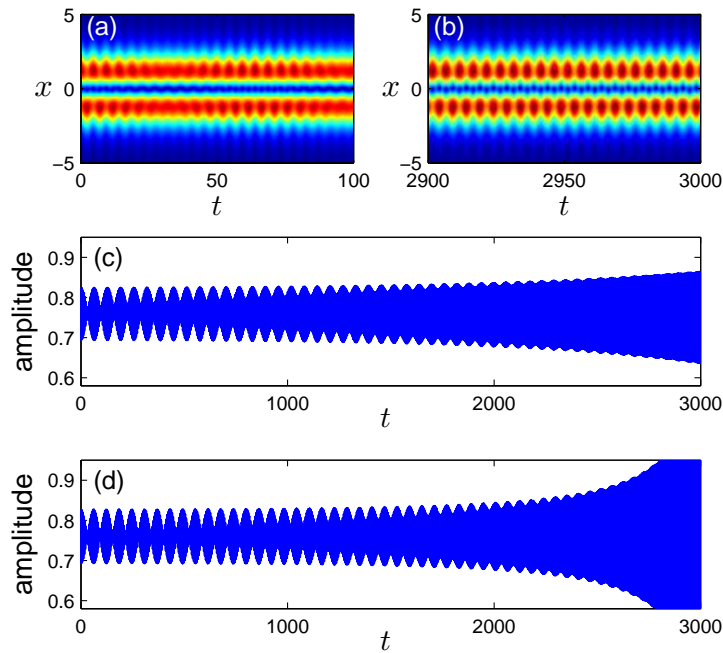}}
\caption{(Color online) Solution evolution below Hamiltonian-Hopf bifurcation in Example 1. (a,b) Contour plots of the PDE solution in the $(x,t)$ plane
on smaller and larger time intervals; (c) amplitude evolution of the PDE solution versus time, where the amplitude is measured as $|U(x,t)|$
at location $x=1.2$; (d) amplitude evolution of the analytically reconstructed perturbation solution (\ref{e:Uexpansion}).
In (c,d), seemingly solid blue regions actually comprise of very fast oscillations as in Fig.~\ref{f:evolution1}(b,d). } \label{f:evolution2}
\end{figure}

It should be noticed that even though solitons in Example 1 break up
both below and above the Hamiltonian-Hopf bifurcation, this breakup
occurs much more quickly above the bifurcation point since the
soliton in this case is linearly unstable.

\textbf{Example 2}\ Our second example pertains to the case where at the Hamiltonian-Hopf bifurcation point, $2\omega < \mu_0$, hence $\gamma$ is real. In this example, we take
\[ \label{e:Vexample2}
V(x)=-3\left[\mbox{sech}^2(x+1.25)+\mbox{sech}^2(x-1.25)\right],
\]
which is a slightly further-separated double-well potential [shown
in Fig.~\ref{f:example2a}(a)], and $\sigma=1$ (focusing
nonlinearity). This linear potential also admits three discrete
eigenvalues, and the power curve of dipole-type solitons bifurcated
from the middle eigenmode is plotted in Fig.~\ref{f:example2a}(b).
At the propagation constant $\mu_0\approx 3.5562$, a
Hamiltonian-Hopf bifurcation occurs, and the coalesced eigenvalues
on the imaginary axis are $i\omega$, where $\omega \approx 1.5559$.
Notice that $2\omega < \mu_0$, thus resonance does not occur in the
$U_2$ solution and $\gamma$ is real in the normal form
(\ref{e:mu1T})-(\ref{e:BTT}). Specifically the constants in the
normal form for this second example are
\[\alpha\approx 0.0372, \quad \beta\approx 0.0263, \quad \gamma\approx 0.2763,
\]
where the eigenfunction $[\psi_1, \psi_2]^T$ is normalized to have
unit maximum in $\psi_1^2+\psi_2^2$. In this case,
$\widehat{\gamma}=\gamma-\alpha\beta>0$, thus solutions of the
normal form (\ref{e:BTT2}) are periodic (see Fig.~\ref{f:phase}).

At the propagation constant $\mu=\mu_0+0.05$ [marked by a red dot in Fig.~\ref{f:example2a}(b)], the linear-stability spectrum of the soliton is displayed in Fig.~\ref{f:example2a}(c). The presence of a quartet of complex eigenvalues in this spectrum signals that this soliton is above the Hamiltonian-Hopf bifurcation point.

\begin{figure}[h!]
\centerline{\includegraphics[width=0.8\textwidth]{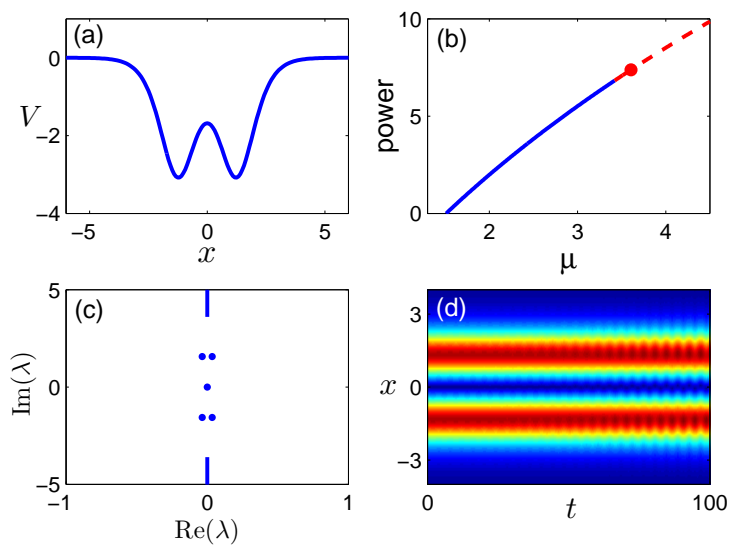}}
\caption{(Color online) (a) The double-well potential (\ref{e:Vexample2}) in Example 2; (b) power curve of dipole-solitons in this potential, where blue indicates stable solitons and red indicates unstable ones; (c) linear-stability spectrum of the unstable soliton marked by a red dot in panel (b); (d) initial evolution of the unstable soliton at the red dot of panel (b) under perturbations. }  \label{f:example2a}
\end{figure}

Now we perturb this unstable soliton by Hopf oscillation mode
(\ref{s:U1}), corresponding to $\epsilon^2\mu_1|_{t=0}=0.05$ and
$\epsilon B|_{t=0}=0.01$ in the perturbation solution
(\ref{e:Uexpansion}). The initial evolution of this perturbed state
is shown in Fig.~\ref{f:example2a}(d), where the instability is seen
to develop. Longer-time amplitude evolution of this perturbed state
is plotted in Fig.~\ref{f:example2b}(a), where the amplitude of the
solution is measured as $|U(x,t)|$ at location $x=1.35$. It is seen
that the envelope of this amplitude evolves quasi-periodically over
a long time, but gradually loses its periodicity in analogy with
that in Fig.~\ref{f:evolution2}(c) for Example 1. Comparatively the
amplitude evolution of analytically reconstructed perturbation
solution (\ref{e:Uexpansion}) is displayed in
Fig.~\ref{f:example2b}(b). The envelope of this analytical amplitude
evolution is periodic, because for this second example,
$\widehat{\gamma}$ is real, $\widehat{\beta}>0$, and
$\widehat{\gamma}>0$, hence the normal form's solutions are periodic
[see Fig.~\ref{f:phase} (upper left panel)]. The numerical and
analytical amplitude evolutions in Fig.~\ref{f:example2b} are in
good agreement over long times (on the order of hundreds of time
units). Their difference over very long times (on the order of
thousands of time units) is due to the fact that for the present
parameters, $3\omega
> \mu_0$, thus resonance with the continuous spectrum will occur in
the $U_3$ solution, and this resonance will break the periodic
oscillation. But our perturbation theory and the resulting normal
form do not account for resonance at such high orders.

\begin{figure}[h!]
\centerline{\includegraphics[width=0.8\textwidth]{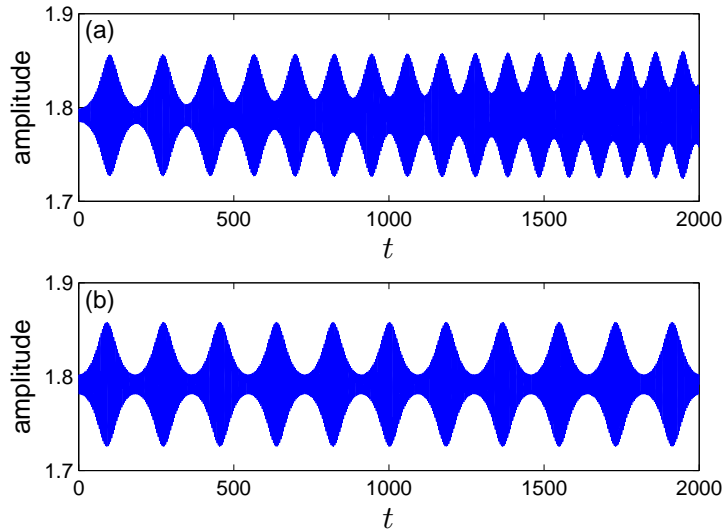}}
\caption{(Color online) (a) Amplitude evolution of the PDE solution for the perturbed soliton in Fig.~\ref{f:example2a}(d) over longer times in Example 2; (b) amplitude evolution of the analytically reconstructed perturbation solution (\ref{e:Uexpansion}). } \label{f:example2b}
\end{figure}

\section{Summary and discussion}

In this paper, we have derived a normal form for general
Hamiltonian-Hopf bifurcations of solitons in NLS equations with
external potentials. This normal form is a simple second-order
nonlinear ODE whose dynamics is analytically predictable, and it is
asymptotically accurate in describing solution dynamics near
Hamiltonian-Hopf bifurcations. We showed that when the nonlinear
coefficient in this normal form is complex, which occurs if the
second harmonic of the Hopf bifurcation frequency is resonant with
the continuous spectrum of the system, the solution of this normal
form blows up to infinity in finite time, indicating that solution
oscillations near Hamiltonian-Hopf bifurcations will strongly
amplify and eventually get destroyed. When the nonlinear coefficient
of the normal form is real, the normal form can admit periodic
solutions, which correspond to long-lasting solution oscillations in
the original PDE system. Quantitative comparison between the normal
form's predictions and true PDE solutions was also made in several
numerical examples, and good agreement was obtained. The normal form
we derived sheds much light on the analytical understanding of
solution behaviors around general Hamiltonian-Hopf bifurcations.

It is noted that even though the normal form in this article was
derived for the specific NLS equation (\ref{e:NLS}) with cubic
nonlinearity and linear potentials, the same calculation can be
trivially extended to generalized NLS equations with arbitrary forms
of nonlinearities and potentials (including nonlinear potentials)
\cite{Yang_classification_2012}. Thus the normal form we derived is
valid for Hamiltonian-Hopf bifurcations in all generalized nonlinear
Schr\"odinger equations.

It is interesting to notice that our normal form can be dissipative
(since its nonlinear coefficient can be complex), even though the
original PDE system (\ref{e:NLS}) is conservative. This dissipative
nature of the normal form is caused by resonant radiation emission
from nonlinearity-induced higher harmonics of oscillating modes.
However, this resonant radiation emission does not lead to the decay
of solution oscillations. Instead, solution oscillations intensity,
as the normal form predicts and the numerics confirms. This
situation is analogous to oscillations induced by an internal mode
with negative Krein signature \cite{Peli_oscillation_2015}.

Lastly, we would like to point out that our normal form does not
admit chaotic solutions, which indicates that solution dynamics near
Hamiltonian-Hopf bifurcations is not chaotic in the PDE system (at
least on the time scale of $\epsilon^{-1}$ for which the normal form
was derived). This analytical prediction is consistent with our
numerical PDE results. In the NLS equation with a triple-well
potential studied in \cite{Goodman_2011}, chaotic motion was
reported in both the ODE model and PDE system. In that case, the
Hamiltonian-Hopf bifurcation occurred at low powers of solitons,
where the chaotic motion was very weak in the ODE model (if at all).
Notice that the ODE model derived in \cite{Goodman_2011} did not
account for resonant radiation emission of oscillating solutions,
thus in the PDE system where such resonant radiation is present,
weak chaotic motion of the ODE model may not materialize. Whether
there is chaotic motion near Hamiltonian-Hopf bifurcations in the
PDE system, on time scales much longer than $O(\epsilon^{-1})$, is a
question which may merit further investigation.

\section*{Acknowledgment}
This work was supported in part by the Air Force Office of
Scientific Research (Grant No. USAF 9550-12-1-0244) and the National
Science Foundation (Grant No. DMS-1311730).


\end{document}